\documentclass[twoside]{article}
\usepackage[paperwidth=6.00in,paperheight=9.00in,margin=1.0in]{geometry}

\usepackage{amsmath, amssymb, graphics}

\newcommand{\mathsym}[1]{{}}

\makeatletter
\long\def\M#1{\leavevmode\setbox\@tempboxa\hbox{#1}\@tempdima\fboxrule
    \advance\@tempdima \fboxsep \advance\@tempdima \dp\@tempboxa
   \hbox{\lower \@tempdima\hbox
  {\vbox{\hrule \@height \fboxrule
          \hbox{  \hskip\fboxsep
          \vbox{\vskip\fboxsep \box\@tempboxa\vskip\fboxsep}\hskip
                 \fboxsep\vrule \@width \fboxrule}%
                  }}}}
\makeatother

\let \ttorg \tt \def \tt{\ttorg \obeyspaces}

\begin{document}

\thispagestyle{empty}

\date{}

\title{}
\leftline{\Large\textrm{THE DIRAC EQUATION AND}}
\vspace{2pt}
\leftline{\Large\textrm{THE MAJORANA DIRAC EQUATION}}
\vspace{4pt}
\hrule

  

\vspace{12pt} 
\begin{center}
{\Large\textsc{Louis H Kauffman}}\\ Department of Mathematics, Statistics and Computer Science \\ 851 South Morgan Street   \\ University of Illinois at Chicago\\
Chicago, Illinois 60607-7045\\ and\\ Department of Mechanics and Mathematics\\ Novosibirsk State University\\Novosibirsk, Russia\\$<$kauffman@uic.edu$>$\\
\end{center} 
\vspace{12pt}
\begin{center}
{\Large\textsc{Peter Rowlands}}\\
Physics Department\\University of Liverpool\\Oliver Lodge Laboratory\\Oxford Street\\Liverpool L69 7ZE\\
$<$p.rowlands@liverpool.ac.uk$>$
\end{center} 

  
\thispagestyle{empty}

\vspace{2cm}

\baselineskip=14pt  

\section{Introduction}
We discuss the structure of the Dirac equation and how the nilpotent and the Majorana operators arise naturally in this context. This provides a link between Kauffman's work on discrete physics, iterants and Majorana Fermions
\cite{KP,KN:QEM,KN:Dirac,NCW, NCWConstraints,MLogic,Iterants,MajoranaBraid}  and the work on nilpotent structures and the Dirac equation of Peter Rowlands \cite{Rowlands1,Rowlands2,Rowlands3,Rowlands4,Rowlands5,Rowlands6}. We give an expression in split quaternions for the  Majorana Dirac equation in one dimension of time and three dimensions of space.\\

\newpage  

In \cite{Majorana} Majorana discovered a version of the Dirac equation that can be expressed entirely over the real numbers. This led him to speculate that the solutions to his version of the Dirac equation would correspond to particles that are their own anti-particles.  It is the purpose of this paper to examine the structure of this Majorana-Dirac Equation, and to find basic solutions to it by using the nilpotent technique. We succeed in this aim and describe our results. The Majorana-Dirac equation can be written as follows:
$$(\partial/\partial t + \hat{\eta} \eta \partial/\partial x + \epsilon \partial/\partial y + \hat{\epsilon} \eta \partial/\partial z - \hat{\epsilon} \hat{\eta} \eta m) \psi = 0$$ where $\eta$ and $\epsilon$ are the simplest generators of iterant algebra with $\eta^{2} = \epsilon^{2} = 1$ and $\eta \epsilon + \epsilon \eta = 0,$
and $\hat{\epsilon}, \hat{\eta}$ form a copy of this algebra that commutes with it. This combination of the simplest Clifford algebra with itself is the underlying structure of Majorana Fermions, forming indeed the underlying structure of all Fermions.  We show how to make nilpotent formulations for Majorana Dirac equations and consequently how to solve these equations via Majorana operators.\\

Here is a concise background about Fermions that will be of use for the rest of this paper. The operator algebra for a Fermion is given by creation and annihilation operators $U^{\dagger}$ and $U$ satisfying the equations $U^{2} =( U^{\dagger})^2 = 0$ and
$UU^{\dagger} + U^{\dagger}U = 1$ \cite{Zee}. Call an algebra generated by $U$ and $U^{\dagger}$  a {\it Fermion algebra} if it satisfies these equations.  For this introduction make the following well-known remark:
Suppose that we are given a Clifford algebra with generators $a$ and $b$ so that $s^2= b^2 = 1$ and $ab + ba = 0.$ It is assumed that $a^{\dagger} = a$ and that 
$b^{\dagger} = b.$ Then we obtain a Fermion algebra from this Clifford algebra by defining $U = (a + i b)/2$ and $U^{\dagger} = (a - ib)/2$ where $i = \sqrt{-1}.$ The reader will have no difficulty verifying this assertion.\\

It has been suggested  \cite{Kitaev,Ivanov} that electrons or other Fermions might behave, under certain circumstances, as if the electron was composed of two particles corresponding to this decomposition into operators $a$ and $b.$ Furthermore, since $a$ and $b$ are invariant under conjugation ($\dagger$), it has been suggested that the particles 
corresponding to $a$ and $b$ are {\it Majorana Fermions}, particles that are their own anti-particles. The reason for this nomenclature goes back to the paper of Majorana
\cite{Majorana} where he constructed a version of the Dirac equation based on real Clifford algebra so the solutions could model particles that were their own anti-particles.
It has been a subject of speculation whether such particles exist. The recent suggestion that electrons themselves are composed of Majorana particles is startling to say the least.
Some experimental evidence is availiable for this hypothesis in terms of the behaviour of electrons in nano-wires \cite{Beenakker,Kou}.  Thus we call the operators $a$ and $b$ the 
{\it Majorana operators} related to the Fermion algebra. Note that $a = (U + U^{\dagger})$ and  $b = (U - U^{\dagger})/i.$\\

While it has been natural to say that the operators $a$ and $b$ are Majorana operators, their relationship to the Majorana Dirac equation has hitherto been obscure.
One purpose of this paper is to show how indeed there are real solutions to the Majorana Dirac equation that are built in terms of the Majorana operators.\\

This paper is organized as follows: Section 2 introduces the Dirac equation, its nilpotent reformulation and the appearance of algebraic Fermion operators as nilpotent algebra elements supporting solutions to the Dirac equation. We explain the formulation of the Majorana-Dirac operator as described above. In Section 3 we use a nilpotent reformulation of the 
Majorana-Dirac operator to find real solutions to the Majorana-Dirac equation and we show how the Clifford algebra of Majorana operators is related to these solutions. In a separate section we give real solutions to the Majorana-Dirac equation specialized to one dimension of space and one dimension of time. We rewrite this specialization in terms of light-cone 
coordinates and compare our results with the Feynman Checkerboard model \cite{Feynman,KN:Dirac}. In Section 4 we reformulate the Dirac equation in terms of {spacetime algebra} by which we mean
a Clifford algebra generated by elements $e_1,e_2,e_3,e_4$ where all pairs of distinct generators anti-commute and the first three generators square to $1$ while the last ($e_4$) 
squares to $-1.$ We prove that the Dirac operator can be written in the form 
$$\partial /\partial t +  e_1 \partial /\partial x  + e_2 \partial /\partial y  + e_3 \partial /\partial z  + e_4 m$$
and that it can be converted to the nilpotent form if and only if there is an element $\mu$ such that $\mu^2 = -1$ and $\mu e_1, \mu e_2, \mu e_3, \mu e_4$ are generators for 
a new spacetime algebra. We then use this result to classify all possible spacetime algebras that can be used to make Majorana-Dirac operators.
In Section 5 we discuss the physics of our results from the point of view of Peter Rowlands. Section 6 is a concluding summary.\\

\section{The Dirac Equation and Majorana\\ Fermions}
 We construct the Dirac equation.  If the speed of light is equal to $1$ (by convention), then energy $E$, momentum $p$ and mass $m$ are
related by the (Einstein) equation $$E^2 = p^2 + m^2.$$ Dirac constructed his equation by looking
for an algebraic square root of $ p^2 + m^2$ so that he could have a linear operator for $E$ that would take the same role as the Hamiltonian in the Schroedinger equation. We will get to this operator by first taking the case where $p$ is a scalar (we use one dimension of space and one dimension of time.).
Let $E = \alpha p + \beta m$ where $\alpha$ and $\beta$ are elements of a possibly non-commutative,
associative algebra. Then $$E^2 = \alpha^2 p^2 + \beta^2 m^2 + pm(\alpha \beta + \beta \alpha).$$
Hence we will satisfiy $E^2 = p^2 +m^2$ if $\alpha^2 = \beta^2 = 1$ and
 $\alpha \beta + \beta \alpha = 0.$ This is a familiar Clifford algebra pattern. Note that this algebra can be represented by $2 \times 2$ matrices with 
 $$\alpha  = \left(\begin{array}{cc}
			-1&0\\
			0&1
			\end{array}\right)$$  and
			 $$\beta  = \left(\begin{array}{cc}
			0&1\\
			1&0
			\end{array}\right).$$ 
			 Then, because the quantum operator for momentum is $-i \partial/\partial x$ and the operator for energy is $i\partial/\partial t,$ we have the Dirac equation $$i\partial \psi /\partial t = -i \alpha \partial \psi /\partial x + \beta m \psi.$$
 Let $${\cal O} = i\partial /\partial t + i \alpha \partial /\partial x - \beta m $$ so that the Dirac equation 
 takes the form $${\cal O} \psi(x,t) = 0.$$ Now note that 
 $${\cal O} e^{i(px - Et)} = (E - \alpha p - \beta m) e^{i(px - Et)}.$$
 We let $$\Delta = (E - \alpha p - \beta m)$$ and let 
 $$ U = \Delta \beta \alpha = (E - \alpha p - \beta m)\beta \alpha = \beta \alpha E + \beta p - \alpha m,$$
 then $$U^2 = -E^2 + p^2 + m^2 = 0.$$ This nilpotent element leads to a (plane wave) solution to the 
 Dirac equation as follows:  We have shown that $${\cal O} \psi = \Delta \psi$$ for 
 $\psi =  e^{i(px - Et)}.$  It then follows that 
 $${\cal O}(\beta \alpha \Delta \beta \alpha \psi) = \Delta \beta \alpha \Delta \beta \alpha  \psi =
 U^2 \psi = 0,$$
  from which it follows that $$\psi = \beta \alpha U e^{i(px - Et)}$$ is a (plane wave) solution to the Dirac equation.\\

In fact, this calculation suggests that we should multiply the operator ${\cal O}$ by $\beta \alpha$ on the right, obtaining the operator 
$${\cal D} = {\cal O} \beta \alpha =  i\beta \alpha \partial /\partial t + i \beta \partial /\partial x - \alpha m, $$
and the equivalent Dirac equation $${\cal D}\psi = 0.$$ In fact for the specific $\psi$ above we will now 
have ${\cal D} (U  e^{i(px - Et)} ) = U^2  e^{i(px - Et)} = 0.$ This idea of reconfiguring the Dirac equation
in relation to nilpotent algebra elements $U$ is due to Peter Rowlands \cite{Rowlands1}.  Rowlands does this in the context of vector (Clifford) and quaternion algebra.
Note that the solution to the Dirac equation that we have found is expressed in Clifford algebra. It can be articulated into specific vector solutions by using a matrix representation of the algebra.
We see that $ U = \beta \alpha E + \beta p - \alpha m$ with $U^2 = 0$ is the essence of  this plane wave solution to the Dirac equation. This means that a natural non-commutative algebra arises directly 
and can be regarded as the essential information in a Fermion. It is natural to compare this algebra structure with algebra of creation and annihilation operators that occur in quantum field theory. To this end we recapitulate and start again in the next subsection.

\subsection{$U$ and $U^{\dagger}$}
We start with $\psi =  e^{i(px - Et)}$ and the operators
$\hat{E} = i \partial/\partial t$ and 
$\hat{p} = -i \partial/\partial x$  so that 
$\hat{E}\psi = E \psi$ and
$\hat{p}\psi = p \psi.$
The Dirac operator is
$${\cal O} = \hat{E} - \alpha \hat{p} - \beta m$$
and the modified Dirac operator is
$${\cal D} ={\cal O} \beta \alpha =  \beta \alpha  \hat{E} + \beta \hat{p} - \alpha m,$$ so that
$$ {\cal D}\psi = (\beta \alpha E + \beta p - \alpha m)\psi = U \psi.$$
If we let $\tilde{\psi} =  e^{i(px +Et)}$ (reversing time), then we have
$$ {\cal D}\tilde{\psi} = (-\beta \alpha E + \beta p - \alpha m)\psi = U^{\dagger} \tilde{\psi},$$
giving a definition of $U^{\dagger}$ corresponding to the anti-particle for $U\psi.$
We have
$$U =  \beta \alpha E + \beta p - \alpha m$$
and
$$U^{\dagger} =  - \beta \alpha E + \beta p - \alpha m$$
Note that here we have 
$$(U + U^{\dagger})^2 =  (2 \beta p + \alpha m)^2 =  4 (p^2 + m^2 )= 4 E^2 ,$$ and
$$(U - U^{\dagger})^2 = - ( 2 \beta \alpha E)^2 = - 4 E^2 .$$
We have that $$U^{2} = (U^{\dagger})^{2} = 0 $$ and $$U U^{\dagger} + U^{\dagger} U = 4 E^{2}.$$
Thus we have a direct appearance of the Fermion algebra corresponding to the Fermion plane wave solutions to the Dirac equation. Furthermore, as shall see below, the decomposition of $U$and $U^{\dagger}$ into the
corresponding Majorana Fermion operators corresponds to $E^2 = p^2 + m^2 .$\\

\noindent To see this, normalize by dividing by $2 E$ we have
$$U = (A + Bi)E$$ and $$U^{\dagger} = (A - Bi)E, $$
with 
$$A =( \beta p + \alpha m)/E $$ and 
$$B = i \beta \alpha.$$ so that
$$A^2 = B^2 = 1$$ and $$AB + BA = 0.$$
This shows how the Fermion operators are expressed in terms of the simpler Clifford algebra of Majorana operators. (See the introduction to this paper for a discussion of the 
 role of Majorana operators.)

\bigbreak

\subsection{Writing in the Full Dirac Algebra}
So far, we have written the Dirac equation in one dimension of space and one dimension of time.
We give here a way to boost the formalism directly to three dimensions of space. We take an independent Clifford algebra generated by $\sigma_{1}, \sigma_{2}, \sigma_{3}$ with
$\sigma_{i}^{2} = 1$ for $i=1,2,3$ and $\sigma_{i}\sigma_{j} = - \sigma_{j}\sigma_{i}$ for 
$i \ne j.$ Now assume that $\alpha$ and $\beta$ as we have used them above generate an independent Clifford algebra that commutes with the algebra of the $\sigma_{i}.$ Replace
the scalar momentum $p$ by a $3$-vector momentum $p = (p_1 , p_2 , p_3 )$ and let 
$p \bullet \sigma = p_{1} \sigma_{1} +  p_{2} \sigma_{2} +  p_{3} \sigma_{3}.$ We replace
$\partial / \partial x$ with $\nabla  = (\partial /  \partial x_{1} , \partial / \partial x_{2}, \partial / \partial x_{2} )$
and $\partial p / \partial x$ with $\nabla \bullet p.$
\bigbreak

\noindent We then have the following form of the Dirac equation.
$$i\partial \psi /\partial t = -i \alpha \nabla \bullet \sigma  \psi  + \beta m \psi.$$
 Let $${\cal O} = i\partial /\partial t + i \alpha \nabla \bullet \sigma  - \beta m $$ so that the Dirac equation 
 takes the form $${\cal O} \psi(x,t) = 0.$$  In analogy to our previous discussion we let 
 $$\psi(x,t) =  e^{i(p \bullet r - Et)}$$  where $p=(p_x,p_y,p_z)$ and $r=(x,y,z)$ and $\bullet$ denotes the dot product. We construct solutions by first applying the Dirac operator to this $\psi.$ The two Clifford algebras interact to generalize directly the nilpotent solutions and Fermion algebra that we have detailed for one spatial dimension to this three dimensional case. To this purpose the modified Dirac operator is
$$ {\cal D} = i\beta \alpha \partial/\partial t + \beta \nabla \bullet \sigma - \alpha m.$$
And we have that $${\cal D}\psi = U \psi$$ where
$$U = \beta \alpha E + \beta p \bullet \sigma - \alpha m.$$
We have that $U^{2}= 0$ and  $U \psi$ is a solution to the modified Dirac Equation, just as before.
And just as before, we can articulate the structure of the Fermion operators and locate the corresponding Majorana Fermion operators. We leave these details to the reader.
 \bigbreak
 
 \subsection{Majorana Fermions}
There is more to do. We will now make a Dirac algebra distinct from the one generated by $\alpha, \beta, \sigma_1 , \sigma_2 , \sigma_3$ to obtain an equation that can have real solutions. This was the strategy that Majorana \cite{Majorana} followed to construct his Majorana Fermions. A real equation can have solutions that are invariant under complex conjugation and so can correspond to particles that are their own anti-particles. We will describe this Majorana algebra in terms of the split quaternions $\epsilon$ and $\eta.$ For convenience we use the matrix representation given below.  
$$ \epsilon = \left(\begin{array}{cc}
			-1&0\\
			 0&1
			\end{array}\right),
			 \eta = \left(\begin{array}{cc}
			0&1\\
			1&0
			\end{array}\right).$$
Let $\hat{\epsilon}$ and $\hat{\eta}$ generate another, independent algebra of 
split quaternions, commuting with the first algebra generated by $\epsilon$ and $\eta.$
Then a totally real Majorana Dirac equation can be written as follows:
$$(\partial/\partial t + \hat{\eta} \eta \partial/\partial x + \epsilon \partial/\partial y + \hat{\epsilon} \eta \partial/\partial z - \hat{\epsilon} \hat{\eta} \eta m) \psi = 0.$$
\bigbreak

\noindent To see that this is a correct  Dirac equation, note that
$$\hat{E} = \alpha_{x} \hat{p_{x}} +  \alpha_{y} \hat{p_{y}} +  \alpha_{z} \hat{p_{z}} + \beta m$$
(Here the ``hats'' denote the quantum differential operators corresponding to the energy and momentum.)
will satisfy $$\hat{E}^{2} =  \hat{p_{x}}^{2} + \hat{p_{y}}^{2} + \hat{p_{z}}^{2} + m^{2}$$ if the algebra
generated by $\alpha_{x}, \alpha_{y}, \alpha_{z}, \beta$ has each generator of square one and each distinct pair of generators anti-commuting. From there we obtain the general Dirac equation by replacing
$\hat{E}$ by $i\partial/\partial t$, and $\hat{p_{x}}$ with $-i\partial/\partial x$ (and same for $y,z$).
$$(i\partial/\partial t +i\alpha_{x}\partial/\partial x +i\alpha_{y} \partial/\partial y +i\alpha_{z} \partial/\partial y - \beta m) \psi = 0.$$ 
This is equivalent to
$$(\partial/\partial t  +\alpha_{x}\partial/\partial x  + \alpha_{y} \partial/\partial y + \alpha_{z} \partial/\partial y +i \beta m) \psi = 0.$$
Thus, here we take $$\alpha_{x} = \hat{\eta} \eta, \alpha_{y} =  \epsilon, \alpha_{z} = \hat{\epsilon} \eta ,
\beta = i\hat{\epsilon} \hat{\eta} \eta,$$ and observe that these elements satisfy the requirements for the Dirac algebra. Note how we have a significant interaction between the commuting square root of minus one ($i$) and the element $\hat{\epsilon} \hat{\eta}$ of square minus one in the split quaternions. This brings us back to our original considerations about the source of the square root of minus one. Both viewpoints combine in the element $\beta = i \hat{\epsilon} \hat{\eta} \eta$ that makes this Majorana algebra work. Since the algebra appearing in the Majorana Dirac operator is constructed entirely from two commuting copies of the split quaternions, there is no appearance of the complex numbers, and when written out in tensor products of $2 \times 2$ matrices we obtain coupled real differential equations to be solved. Clearly this ending is actually a beginning of a new study of Majorana Fermions. 
\bigbreak			
		
\section{Nilpotents and\\ the Majorana-Dirac Equation}

Let ${\cal D} = (\partial/\partial t + \hat{\eta} \eta \partial/\partial x + \epsilon \partial/\partial y + \hat{\epsilon} \eta \partial/\partial z - \hat{\epsilon} \hat{\eta} \eta m). $
In the last section we have shown how  ${\cal D}$ can be taken as the Majorana operator for which we can look for real solutions to the Dirac equation.
Letting $\psi(x,t) =  e^{i(p \bullet r - Et)},$  we have
$${\cal D}\psi =( -iE + i(\hat{\eta} \eta p_{x} + \epsilon p_{y} + {\hat \epsilon} \eta p_{z}) - {\hat \epsilon}{\hat \eta} \eta m)\psi.$$
Let $$\Gamma = ( -iE + i(\hat{\eta} \eta p_{x} + \epsilon p_{y} + {\hat \epsilon} \eta p_{z}) - {\hat \epsilon}{\hat \eta} \eta m)$$ and 
$$U = \epsilon \eta \Gamma =  ( i(- \eta \epsilon E -\hat{\eta} \epsilon p_{x} + \eta p_{y} - \epsilon {\hat \epsilon} p_{z}) + \epsilon {\hat \epsilon} {\hat \eta} m).$$
The element $U$ is nilpotent, $U^2 = 0,$ and we have that
$$U = A + iB,$$
$$AB + BA = 0,$$
$$A =  \epsilon {\hat \epsilon} {\hat \eta} m,$$
$$B=  - \eta \epsilon E -\hat{\eta} \epsilon p_{x} + \eta p_{y} - \epsilon {\hat \epsilon} p_{z},$$
$$A^2 = -m^2,$$ and 
$$B^2 = -E^2 + p_{x}^2 + p_{y}^2 + p_{z}^2 = -m^2.$$
Letting $\nabla = \epsilon \eta {\cal D},$ we have a new Majorana Dirac operator with $\nabla \psi = U \psi$ so that
$\nabla (U\psi) = U^2 \psi = 0.$
Letting $\theta = (p \bullet r - Et),$ we have 
$$U \psi = (A + Bi)e^{i \theta} = (A + Bi)(Cos(\theta) + i Sin(\theta)) =$$
$$(A Cos(\gamma) -BSin(\theta)) + i(B Cos(\theta) + A Sin(\theta)).$$
Thus we have found two real solutions to the Majorana Dirac Equation:
$$\Phi = A Cos(\theta) - B Sin(\theta) $$ and
$$\Psi = B Cos(\theta) + A Sin(\theta)$$ with
$$\theta= (p \bullet r - Et)$$  and $A$ and $B$ the Majorana operators described above.
Note how the Majorana Fermion algebra generated by $A$ and $B$ comes into play in the construction of these solutions.\\

We take it as quite significant that the Majorana algebra is directly involved in these solutions. In other work \cite{MajoranaBraid,MLogic,Iterants,KP} we review the main features of recent applications of the Majorana algebra and its relationships with representations of the braid group and with topological quantum computing. We are  now in a position to assess the relationship of the Majorana algebra with actual solutions to the Majorana-Dirac equation, and this will be the subject of subsequent work.\\

\subsection{Spacetime in $1 + 1$ dimensions.}
Using the method of this section and spacetime with one dimension of space ($x$), we can write a real Majorana Dirac operator in the form
$$\partial/\partial t + \epsilon \partial/\partial x + \epsilon \eta m$$ where, the matrix representation is now two dimensional with 
$$ \epsilon = \left(\begin{array}{cc}
			-1&0\\
			 0&1
			\end{array}\right),
			 \eta = \left(\begin{array}{cc}
			0&1\\
			1&0
			\end{array}\right),
			\epsilon \eta  = \left(\begin{array}{cc}
			0&-1\\
			1&0
			\end{array}\right).$$
We obtain a nilpotent operator, ${\cal D}$ by multiplying by $i \eta:$
$${\cal D} = i \eta \partial/ \partial t + i \eta \epsilon \partial / \partial x - i \epsilon m.$$
Letting $\psi = e^{i(px - Et)},$ we have $${\cal D} \psi = (A + i B)\psi$$ where
$$A = \eta E + \epsilon \eta p$$ and 
$$B = - \epsilon m.$$
Note that $A^2 = E^2 - p^2 = m^2$ and $B^2 = m^2,$ from which it is easy to see that $A + i B$ is nilpotent. $A$ and $B$ are the Majorana operators for this decomposition.
Multiplying out, we find 
$$(A + iB) \psi = (A + i B)(cos(\theta) + i sin( \theta)) = $$
$$(A cos(\theta) - B sin (\theta)) + i( B cos(\theta) + A sin (\theta))$$
where $\theta = px - Et.$
We now examine the real part of this expression, as it will be a real solution to the Dirac equation. The real part is
$$A cos(\theta) - B sin (\theta) = (\eta E + \epsilon \eta p) cos(\theta) + em sin(\theta)$$
$$= \left(\begin{array}{cc}
			-m sin(\theta) & (E-p) cos(\theta)\\
			(E+p) cos(\theta) & m sin(\theta)
			\end{array}\right).$$
Each column vector is a solution to the original Dirac equation corresponding to the operator $$\nabla = \partial/\partial t + \epsilon \partial/\partial x + \epsilon \eta m$$
written as a $2 \times 2$ matrix differential operator. We can see this in an elegant way by changing to light-cone coordinates:
$$ r = \frac{1}{2}(t + x), l =  \frac{1}{2}(t - x).$$ (Recall that we take the speed of light to be equal to $1$ in this discussion.) Then
$$\theta = px - Et = -(E-p)r - (E + p)l.$$ and the Dirac equation
$$(\partial/\partial t + \epsilon \partial/\partial x + \epsilon \eta m)  \left(\begin{array}{c}
			\psi_{1}\\
			\psi_{2}
			\end{array}\right) = 0$$ becomes the pair of equations
$$\partial \psi_{1}/ \partial l = m \psi_{2},$$
$$\partial \psi_{2}/ \partial r= - m \psi_{1}.$$
Note that these equations are satisfied by 
$$\psi_{1} = - m sin(-(E-p)r - (E + p)l),$$ 
$$\psi{_2} = (E+p)cos(-(E-p)r - (E + p)l)$$
exactly when  $E^2 = p^2 + m^2$ as we have assumed.
It is quite interesting to see these direct solutions to the Dirac equation emerge in this $1 + 1$ case. The solutions are fundamental and they are distinct from the 
usual solutions that emerge from the Feynman Checkerbooad Model \cite{Feynman,KN:Dirac}. It is the above equations that form the basis for the Feynman Checkerboard model that is obtained by examining paths in a discrete
Minkowski plane generating a path integral for the Dirac equation. We will investigate the relationship of this approach with the Checkerboard model in
a subsequent paper.\\

\section{Spacetime Algebra}
Another way to put the Dirac equation is to formulate it in terms of a {\it spacetime algebra.} By a spacetime algebra we mean a Clifford algebra with generators
$\{ e_1, e_2, e_3, e_4 \}$ such that $e_{1}^2 = e_{2}^2 = e_{3}^2 = 1$, $e_{4}^2 = -1$ and $e_{i}e_{j} + e_{j}e_{i} = 0$ for $i \ne j.$ Thus the generators of the algebra
fit the Minkowski metric and we can represent a point in space time by $p = xe_1 + ye_2 + ze_3 + te_4$ so that $p^2 = x^2 + y^2 + z^2 - t^2$ corresponds to the spacetime metric 
with the speed of light $c = 1.$ (The reader may wish to compare this approach with Hestenes \cite{Hestenes}.)\\

Since the Dirac algebra demands $\{ \alpha_1, \alpha_2, \alpha_3, \beta \}$ with all  elements squaring to $1$ and anti-commuting, we see that spacetime algebra is interchangeable with Dirac algebra via the translation: $$ \alpha_1 = e_1 , \alpha_2 = e_2 , \alpha_3 = e_3, \beta \ = -i e_4 $$ where $i= \sqrt{-1}$ is a square root of negative unity that commutes with all algebra elements.\\

The standard Dirac equation is $${\cal O} \psi = 0$$ where  
$${\cal O} = i\partial /\partial t + i \alpha_1 \partial /\partial x  +i \alpha_2 \partial /\partial y  +i \alpha_3 \partial /\partial z - \beta m. $$
Thus we can rewrite ${\cal O}$ as
$${\cal O} = i\partial /\partial t + i e_1 \partial /\partial x  +i e_2 \partial /\partial y  +i e_3 \partial /\partial z + i e_4 m. $$
Then, multiply the whole Dirac equation by $-i$ and we find the equivalent operator
$${\cal O'} = \partial /\partial t +  e_1 \partial /\partial x  + e_2 \partial /\partial y  + e_3 \partial /\partial z  + e_4 m. $$
This point of view makes it clear how to search for Majorana algebra since we can search for a spacetime algebra of real matrices. Then the Dirac equation in the form
$${\cal O'} \psi = 0$$ will be an equation over the real numbers. In fact the algebra that we have already written for Majorana is a spacetime algebra:
$$e_1 = \hat{\eta} \eta, e_2=  \epsilon, e_3 = \hat{\epsilon} \eta , e_4 = \hat{\epsilon} \hat{\eta} \eta.$$
Furthermore, we can see that the following lemma gives us a guide to constructing nilpotent formulations of the Dirac equation.\\

\noindent {\bf Definition 1.} Suppose that $\{ e'_1, e'_2, e'_3, e'_4 \}$  generates a spacetime algebra $\cal{A}$ and that $\mu$ is an element of $\cal{A}$ with $\mu^2 = -1$ and  
so that $\{ e_1 = \mu e'_1,  e_2 = \mu e'_2,  e_3 = \mu e'_3,  e_4 = \mu e'_4 \}$ is also a spacetime algebra with $e_{1}^2 = e_{2}^2 = e_{3}^2 = 1$, $e_{4}^2 = -1$ and $e_{i}e_{j} + e_{j}e_{i} = 0$ for $i \ne j.$  Under these circumstances, we call the spacetime algebra  $\cal{A}$ {\it nilpotent}.\\

\noindent {\bf Lemma.} Let $\cal{A}$ be a nilpotent spacetime algebra, with notation as in Definition1  above. Then the operator $${\cal D} = \mu \partial /\partial t +  e_1 \partial /\partial x  + e_2 \partial /\partial y  + e_3 \partial /\partial z  + e_4 m$$ generates
a nilpotent Dirac equation.\\

\noindent {\bf Proof.} We wish to show that if $\psi = e^{i(p \bullet (x,y,z) - Et)}$ and ${\cal D}\psi = U \psi$ then $U^2 = 0.$
Calculating, we find that 
$$U =i( -\mu E + p \bullet (e_1,e_2,e_3)) + e_4 m.$$ It ifollows that $$U^2 = -(  -E^2 + p_{x}^2 + p_{x}^2 + p_{x}^2) - m^2 =   E^2 - p_{x}^2 - p_{y}^2 - p_{z}^2  -m^2 = 0.$$
This completes the proof.  
$\square$\\

\noindent {\bf Example 1.}\\
 Before proceeding to the Majorana structure, consider the standard Dirac algebra. Here we have $\sigma_1,\sigma_2, \sigma_3$ with 
$\sigma_i^{2}= 1$ for each $i = 1,2,3$ and each pair of distinct operators anticommutes. This can be taken to be the Pauli algebra and is represented by matrices over the complex numbers. We take $\alpha$ and $\beta$ as before to generate a Clifford algebra that commutes with the Pauli algebra and is independent of it. Then the associated spacetime algebra has generators $$e'_1 = \alpha \sigma_1, e'_2 = \alpha \sigma_2, e'_3 = \alpha \sigma_3, e'_4 = \sqrt{-1} \beta$$ and the nilpotency corresponds to the fact that these generators, multiplied by $\beta \alpha,$ yield another spacetime algebra. This is given by 
$$e_1 = \mu e'_1 =  \beta \alpha  \alpha \sigma_1 =  \beta \sigma_1 $$
$$e_2 = \mu e'_2 = \beta \alpha   \alpha \sigma_2 =  \beta \sigma_2 $$
$$e_3 = \mu e'_3 = \beta \alpha \alpha \sigma_3  = \beta \sigma_3 $$
$$e_4 = \mu e'_4 = \beta \alpha \sqrt{-1} \beta  =  - \sqrt{-1} \alpha  $$
The corresponding nilpotent Dirac operator is
$${\cal D} = \mu \partial/\partial t + e_1  \partial/\partial x + e_2  \partial/\partial y + e_3  \partial/\partial z    + e_4 m. $$ Hence 
$${\cal D} = \beta \alpha  \partial/\partial t +   \beta \sigma_1  \partial/\partial x +  \beta \sigma_2 \partial/\partial y +  \beta \sigma_3 \partial/\partial z    - \sqrt{-1} \alpha m.$$
Applying this operator to $\psi = e^{\sqrt{-1} (p \bullet r - Et)}$ we obtain the nilpotent
$$A = -  \beta \alpha  \sqrt{-1} E  +   \beta \sigma_1 \sqrt{-1}  p_x +  \beta \sigma_2  \sqrt{-1} p_y +  \beta \sigma_3 \sqrt{-1} p_z    - \sqrt{-1} \alpha m.$$
This can be replaced by the nilpotent 
$$U =  -  \beta \alpha E  +   \beta \sigma_1 p_x +  \beta \sigma_2  p_y +  \beta \sigma_3   p_z   - \alpha m$$
by factoring out the common square root of minus one. This is the same nipotent that we have previously derived. Note that in relation to this standard Dirac algebra we have
the conjugate nilpotent
$$U^{\dagger}=  -  \beta \alpha E  +   \beta \sigma_1 p_x +  \beta \sigma_2  p_y +  \beta \sigma_3   p_z   - \alpha m,$$ and that 
$$U + U^{\dagger} = 2(\beta \sigma_1 p_x +  \beta \sigma_2  p_y +  \beta \sigma_3   p_z   - \alpha m)$$ so that 
$$UU^{\dagger} + U^{\dagger}U = (U + U^{\dagger})^2 = 4(p^2 + m^2) = 4 E^2.$$
This is as we have derived earlier in the paper. The decomposition into Clifford operators follows these lines, giving Clifford elements that square to $E^2.$
When we work with the real spacetime algebras (below) that correspond to the Majorana Dirac equation, the decomposition into Clifford algebras takes a different pattern,
centering on the mass $m$ rather than the energy $E.$\\

\noindent {\bf Example 2.}\\
In the case we have considered with $$e'_1 = \hat{\eta} \eta, e'_2=  \epsilon, e'_3 = \hat{\epsilon} \eta , e'_4 = \hat{\epsilon} \hat{\eta} \eta.
$$ We take 
$\mu = \epsilon \eta$ and we find
$$e_1 = \epsilon \eta \hat{\eta} \eta = \epsilon \hat{\eta}, $$
$$e_2=  \epsilon \eta \epsilon =  - \eta, $$
$$e_3 = \epsilon \eta \hat{\epsilon} \eta =  \epsilon  \hat{\epsilon} ,$$
$$e_4 = \epsilon \eta \hat{\epsilon} \hat{\eta} \eta= \epsilon \hat{\epsilon} \hat{\eta} .$$

\newpage 
Indeed this gives a spacetime algebra and hence a nilpotent Majorana Dirac operator
$${\cal D} = \epsilon \eta  \partial /\partial t +  \epsilon \hat{\eta} \partial /\partial x  - \eta \partial /\partial y  + \epsilon  \hat{\epsilon} \partial /\partial z  + \epsilon \hat{\epsilon} \hat{\eta} m.$$\\

\noindent {\bf Example 3.}\\
Here is another example. We take
$$e'_1 = \hat{\epsilon}, e'_2= \hat{\eta} , e'_3 = \epsilon \eta \hat{\epsilon}\hat{\eta} , e'_4 = \epsilon \hat{\epsilon} \hat{\eta}$$  and 
$\mu = \eta \hat{\epsilon} \hat{\eta}$ and find
$$e_1 = \eta \hat{\epsilon} \hat{\eta}\hat{\epsilon} = -\eta \hat{\eta}, $$
$$e_2=  \eta \hat{\epsilon} \hat{\eta}\hat{\eta} = \eta \hat{\epsilon}, $$
$$e_3 =  \eta \hat{\epsilon} \hat{\eta}\epsilon \eta \hat{\epsilon}\hat{\eta}=  \epsilon,$$
$$e_4 = \eta \hat{\epsilon} \hat{\eta}\epsilon \hat{\epsilon} \hat{\eta} =  -\eta \epsilon.$$
This gives a spacetime algebra and hence a nilpotent Dirac operator
$${\cal D} = \eta \hat{\epsilon} \hat{\eta} \partial /\partial t - \eta \hat{\eta} \partial /\partial x  + \eta \hat{\epsilon} \partial /\partial y   + \epsilon \partial /\partial z  -\eta \epsilon m.$$\\

\noindent {\bf Example 4.} We now give a number of examples of spacetime algebras. For this purpose it is useful to change notation. We will use 
$$I = \epsilon, J = \eta,  i = {\hat  \epsilon}, j = {\hat \eta }.$$ Thus $I^2 = J^2 = i^2 = j^2 = 1$ and $IJ+JI=0$ and $ij+ji=0.$
We will indicate a spacetime algebra as a $4$-tuple $(e_1,e_2,e_3,e_4)$ where we require that the $e_i$ anti-commute and that the squares of the first three $e_i$ are $1$ while
$e_4^{2} = -1.$ The following are spacetime algebras.
$$A = (Jj,I,Ji,Jij)$$
$$B = (Ii,j,Ji,IJi)$$
$$C= (iJ,Ê ÊÊI,Ê ÊÊjJ,ÊÊ ijJ)$$
$$D= (iJ, I,  jJ, IJ)$$ Ê
It is easy to see that $A$,$B$, $C$ and $D$ are nilpotent. Note that (up to signs) $B$ is obtained from $A$ by interchanging $i,j$ with $I,J$ and then interchanging $i$ and $j.$
$C$ is obtained from $A$ by interchanging $i$ and $j$ directly. To see that $A$ is nilpotent, multiply by $IJ.$ The algebra $D$ is also nilpotent, via multiplying by $ijJ.$\\

\noindent {\bf The General Case.} Now suppose that we are given a nilpotent spacetime algebra specified by $\{ e'_1,e'_2,e'_3,e'_4\}$ and $\mu$ with $\mu^2=-1$ so that $\{ e_1,e_2,e_3,e_4\}$ is also 
a spacetime algebra with $e_i = \mu e'_i$ for $i=1,2,3,4.$ Then we have the nilpotent Dirac operator associated with this algebra:
$${\cal D} = \mu \partial/\partial t + e_1  \partial/\partial x + e_2  \partial/\partial y + e_3  \partial/\partial z    + e_4 m.$$
Let $\iota = \sqrt{-1}$, a square root of negative unity that commutes with all algebra elements.
Applying ${\cal D}$ to $\psi = e^{\iota (p \bullet r - Et)}$ we obtain the nilpotent
$$A = \iota (-\mu E + e_1 p_x + e_2 p_x + e_3 p_x) + e_4 m.$$ 
The nilpotent $A$ is directly decomposed into its two (Majorana) Clifford parts as the real and imaginary parts of $A$, just as in our previous discussion of a special case.
Other examples lead to real solutions to the Majorana Dirac equation just as we have done above.
Note that the Clifford parts are 
$$\rho = -\mu E + e_1 p_x + e_2 p_x + e_3 p_x$$
and
$$\tau = e_4 m$$ with 
$\rho^2 = \tau^2 = -m^2$ and $\rho$ and $\tau$ anticommute. It is of interest to note that the Clifford algebra is  collapsed when the mass is equal to zero.\\

But we need to be systematic here. Consider that the fourth elememt of a spacetime algebra has square $-1.$ Up to symmetries the possibilities are $ij$ and $IJi.$
Take each of these cases in turn. First suppose that $e_4 = ij.$ Then consider first all square one elements. These are
$$S = \{i , j, I, J, ijIJ, iI, iJ, jI, jJ \}.$$
The subset of elements of $S$ that anti-commute with $ij$ is 
$$S[ij] = \{ i, j, iI, iJ, jI, jJ\},$$ and the (up to order and symmetry) the only triplet in $S[ij]$ that mutually anti-commutes is 
$$\{ i, jI, jJ\}.$$
This gives the spacetime algebra
$$\{i,jI, jJ, ij\}.$$
This algebra is nipotent via multiplication by $IJi.$\\

Now consider the subset of elements of $S$ that anti-commute with $IJi.$
This subset is 
$$S[IJi] = \{ j, I, J, ijIJ, iI, iJ\}.$$
The triplets that anti-commute are
$$\{ j,iI, iJ\}$$ and
$$\{ijIJ, I, J\}.$$
These give rise to spacetime algebras
$$\{ j,iI,iJ,IJi\}$$
and $$\{ijIJ, I, J, IJi\}.$$
The first is nilpotent via the multiplier $ij$ and the second is nilpotent via the multiplier $IJj.$
Up to symmetries these are all the cases and so we have proved the result    
\newpage 
\noindent {\bf Theorem.} {\it All real Majorana spacetime algebras are nilpotent and, up to permutations and substitutions, they are of the following types:}
$$\{i,jI, jJ, ij\},$$
$$\{ j,iI,iJ,IJi\},$$
$$\{ijIJ, I, J, IJi\}.$$\\

\noindent In a subsequent paper we shall follow up the consequences of this result.\\

\vspace{12pt}    
 \section{Comparison With Rowlands' Viewpoint}
To compare the Majorana equation with the Rowlands nilpotent formulation, we take the four algebraic operators $\epsilon$, $\eta$ and  ${\hat \epsilon}$, ${\hat \eta}$
as components of a double vector set $I$, $J$ and $i, j$.  That is, we are given that $I^2 = J^2 = 1$ and that $IJ + JI = 0$ and we are given that $i^2 = j^2 =1$ and that 
$ij + ji = 0.$ Furthermore, the elements $i$ and $j$ commute with the elements $I$ and $J.$ For translation, we can set 
$$\epsilon = I, \eta = J,  {\hat  \epsilon} = i, {\hat \eta }= j.$$
We use $\iota$ for a commuting square root of negative unity.
Note that while we use $i,j,I,J$ for these algebras, all these elements square to one.
We can then write the Majorana equation in the form.
$$(\partial/\partial t + I i \partial/\partial x + j \partial/\partial y + Ji\partial/\partial z - JIim )\psi = 0.$$
Note that $\{ e_1 = Ii, e_2 = j, e_3 = Ji, e_4 = JIi \}$ forms a spacetime algebra, and that if we multiply each element by $ij$ we get a new spacetime algebra:
$$\{ ij e_1 = -jI, ije_2 = i, ije_3 = -jJ, ije_4 = -jJI \} = $$
$$\{ e'_1 =-jI, e'_2 =i, e'_3 = -jJ, e'_4 = -jJI \}.$$  Thus by our previous discussion this will give a nilpotent formulation of a version of
the Majorana Dirac equation.
Multiplying from the left by $\iota ij$ gives
$$(\iota i j \partial/\partial t + \iota Ij \partial/\partial x - \iota i\partial/\partial y + \iota Jj \partial/\partial z  + \iota IJjm )\psi = 0.$$
Rearranging the symbols leads to
$$(\iota i j \partial/\partial t + \iota jI \partial/\partial x - \iota i\partial/\partial y + \iota jJ \partial/\partial z  + \iota jIJ m )\psi = 0.$$
To compare with the Rowlands Dirac nilpotent formalism, we apply the operator in the bracket to a free particle wavefunction 
$$\psi = e^{\iota(p \bullet  r- Et)}$$ to find  ${\cal D} \psi = A \psi$ with 
$$A = ijE - jIp_x + i p_y - jJp_z + \iota jIJ m.$$
This squares to 
$$-E^2 + p^2 + m^2 = 0.$$
and so is nilpotent. Generally, in the discussion below we will designate such nilpotent factors by the letter $A.$\\

\noindent If we compare this to the Rowlands formalism, where (again recognizing the arbitrary nature of the signs, and also the choice of symbols between $i$, $j$ and $I$, $J$) $A$ can be written in the form
$$A = ij E + jIp_x + \iota jIJp_y + jJp_z + i m.$$
we see that the effect of changing from Dirac to Majorana is to switch the status of the terms $p_y$ and $m.$ In effect, the space-momentum operator no longer has perfect rotation symmetry between its components. This could be related to the fact that experiments so far which have claimed to have detected Majorana behaviour seem to involve only $1-$ or $2-$ dimensional systems. At the same time the Majorana formalism seems to be suggesting a ``mingling" of momentum or angular momentum with mass. If the neutrino is a Majorana particle, then its mass could be generated by a direct exchange of some kind with angular momentum or spin.\\

Here is an alternative representation
$$(\partial/\partial t + I \partial/\partial x + J \partial/\partial y  -ijIJ \partial/\partial z - iIJ m )\psi = 0.$$
Multiply from the left by $ jIJ$:
$$(jIJ \partial/\partial t - jJ  \partial/\partial x  +  jI\partial/\partial y   +  i \partial/\partial z  -  j i m )\psi = 0.$$
Here the third component of momentum has exchanged with the energy term, up to complex factor (rather than the second with mass).
Now apply a free particle wavefunction $\psi = e^{\iota(p \bullet r- Et)}$ to find
$$A = \iota(-jIJE -jJp_x + jI p_y + i p_z) + ij m$$
This is still nilpotent. 
We may note that the same double space structure of $5 + 3$ occurs as in the standard nilpotents. 
We will make a more detailed discussion in a subsequent paper.\\

\subsection{\bf Discrete Dirac}
The key text for Rowland's discrete version of nilpotent Dirac is ``Zero to Infinity" \cite{Rowlands1}, pages 182-184. ``The Foundations of Physical Law" \cite{Rowlands2}  has a more abbreviated version. 
For a formal creation operator that doesn't distinguish between particle and antiparticle mathematically, and can be split into two parts, we could look at the discrete Dirac equation as Rowlands has been writing it, using Kauffman's non-commutative discrete derivatives \cite{KN:QEM, NCW,NCWConstraints}.
Rowlands' discrete version of the nilpotent Dirac equation is of the form below with the option of premultiplying by $\iota$ (a commuting square root of negative unity).
$$(\pm ij \partial/\partial t \pm i\nabla)  (\pm ij E \pm iI P_1 \pm iI P_2 \pm iI P_3 + jm).$$
This means that the creation operator, which automatically generates the nilpotent amplitude is the first bracket with no $m$ term. This can be split into two parts which are negatives of each other, and so could represent particle and antiparticle. But, with no mass term in the operator, the signs could be reversed arbitrarily by premultiplying from the left by $-1.$ So this operator in this form doesn't distinguish particle and antiparticle.\\

Rowlands sees Fermions as the only particles that have a total nonzero weak charge. The weak charge is the only one associated with chirality. He thinks of weak charges as being associated with the pseudoscalar $\iota$ and having an ambiguity with regard to sign,  (this may be related to the discrete time in the models of Kauffman/Noyes \cite{KN:Dirac}  and Garnet Ord \cite{Ord})  and so being ultimately responsible for zitterbewegung, with a weak charge acting like a dipole with its vacuum reflection. The most interesting fermion in this regard is the neutrino, which only has a weak charge. Ambiguity over the sign of the weak charge ($0$ or $2w$ or $0$ or $-2w$ rather than just $0$) appears in those mesons whose decay involves $CP$ violation. In some sense, the zitterbewegung is mixing particle and antiparticle and we are interested in how the Majorana representation can be used here. We are also interested in how the $(1 - \gamma^{5}) /2$ projection operator can be used to create two helicity states when the creation operator for the Fermion using discrete differentials manages to eliminate the mass term.\\

\subsection{\bf On neutrino masses}
This is indirectly related to Majorana. The question is: why does the neutrino have a mass if it remains chiral as to the weak interaction, the only one of the gauge interactions to which it is subject? The nilpotent has the following terms:


$$(ijE + iIp_x + iJp_y + \iota i I J p_z + jm)$$
The charges which relate to these terms are: weak ($ijE$),              strong ($iP$),             electric ($jm$).
Hypothetically massless Fermions are known to have two sharply defined helicity states. If the nilpotent
$$(ijE + iIp_x + iJp_y + \iota iIJp_z + jm)$$
becomes massless, say
$$(ijE + iIp_x + iJ p_y + \iota iIJp_z)$$
which we can write as 
$$(ijE + i P)$$
then chirality is determined by Pauli exclusion because
$$(-ijE + iP)$$
can exist, but not
$$(-ijE-iP)$$
or
$$(ijE - i P).$$

Helicity is taken as the relative sign of $ijE / iP.$ Positive is left-handed, negative is right-handed.
So we have $LH$ Fermions and $RH$ anti-Fermions.\\

How do real Fermions get their masses? We take the three gauge interactions separately. Let's take
$$(ijE + iP + jm)$$
and consider the strong interaction. In Rowlands' representation of baryons, three have $+P$ and three have $-P$  because each direction of $P,$ (the directions are $x,y,z$) introduces a separate
nilpotent bracket and the active component switches between the three directions.
The mass comes from this switching, equivalent to the parity operator (${\cal P}$), with maximum achirality. The $P$ term is in the same position as the strong charge. The two signs of $P$ ensure a large mass for any baryon.\\

How does $(ijE+iP+jm)$ become $(ijE-iP+jm)$   in such a case? We annihilate $(ijE+iP+jm)$  by creating $(-ijE-iP+jm) (LH)$ and at the same time create the new state $(ijE-iP+jm) (RH).$ Thus, to switch between the two spin states requires the exchange of a spin $1$ boson:
$$(-ijE -iP + im)(ijE -iP + jm)$$
The whole process then becomes
$$(ijE + iP + jm)(-ijE -iP + jm)(ijE - iP + jm) \longrightarrow (ijE-iP + jm)$$
The first bracket is the Fermion state to be changed; the next two brackets form the spin 1 boson absorbed, and the bracket on the RHS becomes the final Fermion state. The reverse process would be
$$(ijE-iP+jm)(-ijE+iP+jm)(ijE-iP+jm) \longrightarrow (ijE + iP +jm)$$
with the spin 1 boson becoming
$$(ijE + iP + jm)(-ijE + iP +jm).$$
Of course, the bosons in this case (gluons) are massless but we can leave out the mass if we use the discrete representation of the operator. (A spin $0$ boson, notably, unlike spin $1$, does not change a Fermionic state.)\\

Electrons are Fermions with electric charges, and this interaction is both $LH$ and $RH$, unlike the weak, so electrons have $RH$ states which experience electric but not weak interactions. In Rowlands' view, the electric charge is in the same position as the $jm$ term. We can't switch $jm$ but we can switch $ijE$ and $iP$ simultaneously via charge conjugation ($C$). This doesn't happen directly because the boson required would be$ (-ijE-iP+jm)(-ijE-iP+jm)$, which zeros immediately. So, it would have to be a combination of separate $E$ and $P$ switches. Does mass happen in some way through this process? The two $P$ states give mass immediately. Certainly this is one of the ways mass will be generated, probably the main way. If there is any switching between $+ E$ and $Ð E$, it will be at the level of zitterbewegung (see below).\\

The weak charge is in the same position as the $ijE$ term. To switch this alone is the same operation as time reversal ($T$). If the weak charge is a dipole with its vacuum reflection, then we can consider a zitterbewegung taking place, possibly generating a weak mass. Here, there is no change in $iP.$ The Fermion state is
$$(ijE+iP+jm)$$
and the vacuum state is
$$(-ijE+iP+jm).$$

The zitterbewegung disappears on fixing a real state of the neutrino by observation. If the neutrino mass is due to zitterbewegung it is possible that the mass is observed without two states of spin.
The process of changing the $ijE$ term only requires interaction with a paired Fermion or paired anti-Fermion spin 0 state, such as $(-ijE-iP+jm)(-ijE+iP+jm).$
Applying this will change $(ijE+iP+jm)$ to $(-ijE+iP+jm)$ via
$$(ijE+iP+jm)(-ijE-iP+jm)(-ijE+iP+jm) \longrightarrow (-ijE+iP+jm).$$
The reverse procedure will be
$$(-ikE+ip+jm)(ikE-ip+jm)(ikE+ip+jm) \longrightarrow (-ikE+ip+jm).$$
If a small weak mass is a product of zitterbewegung which is not directly observed in the neutrino state, then maybe the chirality can exist at the same time as mass. Does this mean that the weak mass is not generated in the same way as the others, and so doesn't give us a problem with chirality?\\

The heuristic way of saying a massive particle can't be one-handed is to say that, if a particle has a mass and so can't travel at the speed of light, we could theoretically overtake it and look back and see that it had flipped to the opposite handedness. If the zitterbewegung mass is smaller than the mass from switching parity, then we would not be able to do this since we couldn't find a particle to travel faster than the neutrino. Maybe overtaking the neutrino would be the same as making it its antiparticle (cf Majorana). It would be interesting to see this in relation to neutrino mixing where there are three neutrinos with slightly different masses flipping into each other.\\

We would like to relate neutrino oscillation and zitterbewegung in further work. It is notable that there are three possible switches (see the Theorem at the end of Section 4) between the Majorana and standard Dirac nilpotent algebras. One of these switches the algebra term  of component of momentum with the mass term. Another one switches the algebra
component with the energy term. The third has the appearance of switching (again after multiplication by $\iota$) the energy term and one of the momentum terms from the classical Dirac operator. Perhaps these three modes of switching are analogous to CPT, and responsible for the three neutrino generations.\\

\section{Conclusion}
 We have seen how the nilpotent approach to the Dirac equation sheds new light on the Majorana-Dirac equation and on the structure of Majorana Fermions.
 This paper marks the beginning of new work on this subject. There is much that remains to be done and we will consider the key questions in subsequent papers.
 The direct appearance of Majorana operators (often identified with Majorana Fermions in recent literature) in our real solutions to the Majorana Dirac equation suggests
 a deeper examination of the nature of Majorana Fermions in condensed matter physics and in relation to quantum computing. Our solutions to the Dirac equation in one dimension 
 of space and one dimension of time suggests that it will be useful to re-examine the Feynman Checkerboard model for the Dirac propagator. This should lead to new insight into
 path integral formulations for solutions to the Dirac equation in three dimensions of space and one dimension of time. 
 The most significant possibility for Majorana Fermions outside of condensed matter is in the neutrino sector, where there is a major problem in reconciling chirality with nonzero neutrino masses. A tentative proposal is made here, in the last section, toward a possible resolution.\\

\newpage 
\baselineskip=12pt 

\end{document}